\documentclass[aip,reprint]{revtex4-1}
\usepackage{graphicx}
\usepackage{dcolumn}
\usepackage{bm}
\usepackage[mathlines]{lineno}

\usepackage[utf8]{inputenc}
\usepackage[T1]{fontenc}
\usepackage{mathptmx}
\usepackage{etoolbox}
\usepackage{siunitx}
\usepackage{hyperref}
\usepackage[version=4]{mhchem}
\usepackage{hyperref}

\usepackage{xcolor}

\draft 
\newcommand{\RomanNumeralCaps}[1]{\MakeUppercase{\romannumeral #1}}
\begin{document}


\title{High dynamic-range and portable magnetometer using ensemble nitrogen-vacancy centers in diamond} 



\author{Himanshu Kumar}
\email[]{himanshuk@iitb.ac.in}
\affiliation{Department of Electrical Engineering, Indian Institute of Technology Bombay, Mumbai, India}

\author{Dasika Shishir}
\affiliation{Department of Electrical Engineering, Indian Institute of Technology Bombay, Mumbai, India}

\author{Maheshwar Mangat}
\affiliation{Department of Electrical Engineering, Indian Institute of Technology Bombay, Mumbai, India}

\author{Siddharth Tallur}
\affiliation{Department of Electrical Engineering, Indian Institute of Technology Bombay, Mumbai, India}
\author{Kasturi Saha}
\email[]{kasturis@ee.iitb.ac.in}
\homepage[]{https://www.ee.iitb.ac.in/web/people/kasturi-saha/}
\affiliation{Department of Electrical Engineering, Indian Institute of Technology Bombay, Mumbai, India}
\affiliation{Center of Excellence in Quantum Information, Computing Science and Technology, Indian Institute of Technology Bombay, Mumbai, India}
\affiliation{Center of Excellence Semiconductor Technologies (SemiX), Indian Institute of Technology Bombay, Mumbai, India}

\date{\today}

\begin{abstract}
Nitrogen vacancy (NV) centers in diamonds have been explored for realizing a wide range of sensing applications in the last decade due to their unique quantum properties. Here we realize a compact and portable magnetometer with an ensemble of NV centers which we call the Quantum MagPI (Quantum Magnetometer with Proportional Integral control). Including the sensor head and associated electronics, our sensor assembly can  fit inside \qtyproduct{10x10x7}{\cm} box and control electronics in \qtyproduct{30x25x5}{\cm} box.
We achieve a bandwidth normalized sensitivity of $\sim$ \SI{10}{\nano \tesla \per \sqrt \hertz}. Using closed-loop feedback for locking to the resonance frequency, we extend the linear dynamic range to \SI{200}{\micro \tesla} ($20\times$ improvement compared to the intrinsic dynamic range) without compromising the sensitivity.
We report a detailed performance analysis of the magnetometer through measurements of noise spectra, Allan deviation, and  tracking of  nT-level magnetic fields in real-time. Additionally, we demonstrate the utility of such a magnetometer by real-time tracking the movement of the elevator car and door opening by measuring the projection of the magnetic field along one of the NV-axes under ambient temperature and humidity. 

\end{abstract}


\maketitle 

Quantum sensors based on negatively charged nitrogen vacancy centers in diamond (\ce{NV-}) centers have been extensively used to sense magnetic fields relevant to various scientific and industrial applications such as monitoring large currents in a bus bar of an electric 
vehicle\,\cite{Hatano2022,KUBOTA2023109853} and sensing magnetic fields in  high pressure anvil cells\,\cite{PhysRevB.107.L220102}. Such applications are enabled by the viability of \ce{NV-} centers for wide operating temperature range from sub-\SI{1}{\kelvin} to \SI{1000}{\kelvin}\,\cite{Liu2019} and pressures up to \SI{130}{\giga \pascal}\,\cite{PhysRevB.107.L220102}.
Similarly, high sensitivities achievable with \ce{NV-}
of up to \SI{1}{\pico \tesla} 
\,\cite{PhysRevApplied.10.034044,PhysRevResearch.2.023394} has made it possible to detect very weak magnetic 
fields emanating from very small currents in biological samples\,
\cite{Arai2022,doi:10.1073/pnas.1601513113,Hansen2023}. However, more widespread use of the \ce{NV-} center
depends heavily on the portability and compactness of the sensor. The \ce{NV-} sensor size depends
on various factors like the laser, microwave power used for excitation, bias magnet arrangement, the properties of the diamond
used, and finally the fluorescence collection optics and the sensing electronics. In general, the parameters
that lead to better sensitivity also increase the size of the sensor. For example, the sensitivity of the 
\ce{NV-} based sensor improves by a factor of $\sqrt{P}$, where $P$ is the excitation laser power. Hence,
a laser with higher power can improve the sensitivity. However, such lasers are bulky. Similarly, collection efficiency
can be improved by using optics of higher numerical aperture, which in-turn increase the size of the sensor.  
In the past, extensive work has been 
done to make \ce{NV-} centers as portable and compact as possible, while still maintaining the high 
sensitivity of the sensor. In Ref.\,\cite{Wang22}, Liu.\,\textit{et. al.} obtained a magnetic field 
sensitivity of \SI{21}{\nano \tesla \per \sqrt{\hertz}} on a sensor head of \qtyproduct{4x4x3}{\cm} 
dimensions. Sturner\,\textit{et. al.} in Ref.\,\cite{https://doi.org/10.1002/qute.202000111} has  reported a 
sensor head size of 
\qtyproduct{3x3x2}{\cm} while maintaining a sensitivity of \SI{0.34}{\nano \tesla \per \sqrt{\hertz}}. This is
the most sensitive portable \ce{NV-} magnetometer ever. 
The smallest \ce{NV-} centered has been reported by Donggyu Kim \textit{et. al.}\,\cite{Kim2019}, where, the 
microwave source, photo 
detector, and optical filters have been integrated into a very small area of \qtyproduct{200x200}{\mu \meter} 
using CMOS technology. 
However, the reported sensitivity is only \SI{32}{\micro \tesla \per \sqrt{\hertz}} due the extremely small size of 
the sensing elements. Another approach to miniaturize the sensor is to place the diamond on a fiber head, 
which imparts flexibility to the sensor head. This approach has been taken in 
Refs.,\,\cite{PhysRevApplied.14.044058,Zheng_2020,Kuwahata2020} with the best reported sensitivity of
\SI{0.31}{\nano \tesla \per \sqrt{\hertz}} in Ref.\,\cite{PhysRevApplied.14.044058}.

Further, the use of \ce{NV-} centers in portable applications, especially field applications such as geological sensing or oil and gas sensing, is limited
by the magnetic dynamic range of the sensor. The dynamic range of an \ce{NV-} center in an open-loop 
configuration is approximately $0.1 \Gamma / \gamma_e$\,\cite{10.1063/5.0138301}, where $\Gamma$ is the 
\ce{NV-} line-width of the optically detected magnetic resonance spectrum, and $\gamma_e$ is the electron 
gyromagnetic ratio. On the other hand, the sensitivity of the sensor improves by a factor of $1/ \Gamma$. This
implies that it is desirable to have lower linewidths for better sensitivity, however a lower linewidth 
compromises the dynamic range of the sensor. For example, a typical linewidth of \SI{1}{\mega \hertz} has 
a dynamic range of only around \SI{3.5}{\micro \tesla}. Resonance shift due to temperature shifts in \ce{NV-} correspond to \SI{2.6}{\micro \tesla \per \kelvin}. Therefore, a change in temperature of \SI{1.3}
{\kelvin} can put the \ce{NV-} sensor out of it's dynamic range in the open loop configuration in addition to the effects of any stray fluctuating external magnetic fields. In this work we report a portable, feedback stabilized \ce{NV-} sensor (dubbed as Quantum MagPI) with a sensor head size of \qtyproduct{10x10x7}{\cm}, a sensitivity of \SI{10}{\nano \tesla \per \sqrt{\hertz}} and a dynamic range of \SI{200}{\micro\tesla}, with all electronics enclosed in a standard rack mountable box that can operate in ambient conditions. Our sensor is resistant to changes in temperature and other non-magnetic disturbances because of close-loop feedback tracking.

The \ce{NV-} center is a color defect in the diamond crystal, formed when one of the carbon atoms is replaced by nitrogen and the adjacent atom is knocked out from its lattice site. Negatively charged NV centers (\ce{NV-}) are unique because their ground and excited state are 
spin-triplet states ($S=1$) and their quantum state can be initialized and manipulated via optical pumping 
and microwave (MW) field\,\cite{Rondin_2014}. The ground state of \ce{NV-} is a spin-1 ($S=1$) system, with $m_s=0$ and $m_s=\pm1$ separated 
by a zero-field splitting of $\sim 2.87$ GHz. When pumped with \SI{532}{\nm} green laser, \ce{NV-} goes to 
excited state, emitting red fluorescence (\SI{637}{\nm}) while radiatively decaying to ground state. These transitions are spin conserving. Also, there exists an alternate non-radiative 
decay path to ground state, transferring the spin population from $m_s=\pm1$ to $m_s=0$ state, which enables optical polarization of the \ce{NV-} to $m_s=0$ state. A MW field enables state transfer of the population from $m_s=0$ to $m_s=\pm1$ at resonance which can be mapped by measuring the photo-luminescence (PL) intensity also known as optically detected magnetic resonance\,\cite{Rondin_2014}. With the application of a magnetic field, the 
degenerate states $m_s=\pm1$ are split by Zeeman splitting, which leads to a change in the resonance 
frequency of $m_s = 0$ to $+1$ or $-1$ transition. In addition, the interaction of electron spin with nitrogen 
nuclear spin leads to further splitting into two ($^{15}$N) or three ($^{14}$N) hyper-fine levels. In this 
work, we have utilized a chemical vapor deposition (CVD) grown {100} orientation diamond sample DNV-
B1$^\mathrm{TM}$ of size \qtyproduct{3x3x0.5}{\mm} by Element-6, $\sim1.1$ ppm $^{13}C$ concentration, substitutional nitrogen concentration 
([$N_s^0$]) $\sim 0.8$ ppm, [$NV^-$] concentration $\sim0.3$ ppm.\\

\begin{figure}
 \includegraphics[width=\linewidth]{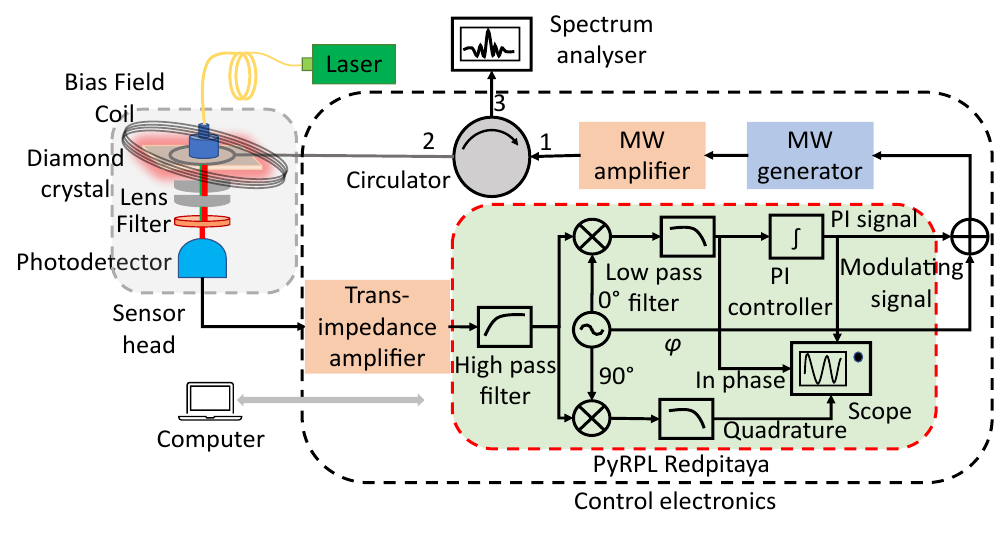}%
 \caption{\label{fig: Schematic} Schematic of the experimental 
 setup.}%
 \end{figure}

A \SI{525}{\nm}, \SI{1.2}{\watt} compact multimode fiber-coupled laser (FL-525-1200) from Lasertack GmbH is used for optical excitation. The output beam from the fiber is collimated with a fixed focus collimator and the 
fiber collimator is mounted on \SI{30}{\mm} cage XY translator (CXY1QA) with the help of mounting adapter (AD11F). 
The diamond is glued on an omega loop antenna printed on a thin substrate (Rogers RO4350B\textsuperscript{TM}). Two 
\SI{25}{\mm} lenses are used to focus the light emitted from the diamond, followed by a \SI{600}{\nm} long 
pass filter (AT600lp) to reject the green light. Finally, the red fluorescence is collected onto Thorlabs photo-detector (DET36A2). The complete assembly is cage-mounted to form a compact sensor head. A hundred-turn current-carrying coil (\SI{7}{\cm} radius) is used to apply a homogeneous bias field. To further improve the footprint of the device, a $3\times 3$  array of $\SI{3}{\cm} \times \SI{3}{\cm}$ permanent magnets (Neodymium $\SI{1}{\cm} \times \SI{1}{\cm} \times \SI{2}{\mm}$) is symmetrically glued on both sides of the sensor head.\\
The electron spin resonances are driven by MW generated using a compact MW synthesizer (Windfreak 
SynthHD(v2)), amplified by three 8.8 dB low noise compact Minicircuits MW amplifiers (ZX60-P33ULN+) 
connected in cascade to provide a gain of 26.8 dB followed by a circulator (PE8432) to prevent damage 
due to power reflection from the antenna. The MW carrier signal is frequency-modulated (FM) by the modulation signal applied at the analog trigger 
signal, configured to generate the FM signal with a Python interface. The FM signal frequency is given by
$f=f_c+f_{dev}\sin(2\pi f_mt)$, where, $f_c$ is the MW carrier frequency, $f_{dev}$ maximum frequency deviation, and $f_m$ is the modulation rate.\\
The experimental control is implemented with the Redpitaya STEMlab 125-14 FPGA board as illustrated in Fig. 
\ref{fig: Schematic}, and an open-source library (PyRPL) is  utilized to perform lockin homodyne demodulation, data acquisition, and closed loop tracking of NV resonances. The fluorescence signal from the photo-detector is fed to the trans-impedance amplifier (Edmund 
optics $\#59-178$) and measured by the Redpitaya board. A high-pass filter is applied to the signal to remove 
the DC offset. The demodulated quadratures are readout using the Redpitaya scope modules, and data is transferred to the computer over the ethernet interface. For closed-loop tracking of the resonance frequency, the phase between 
modulation and demodulation is adjusted such that the entire signal is in one of the quadrature. The signal 
quadrature is input to a proportional-integral (PI) controller, which generates an error signal to lock at 
the resonance frequency. The output of PI controller is added with an analog adder to shift the DC level of 
the modulating signal, which in turn shifts the MW frequency.\\

\begin{figure}
 \includegraphics[width=0.5\textwidth]{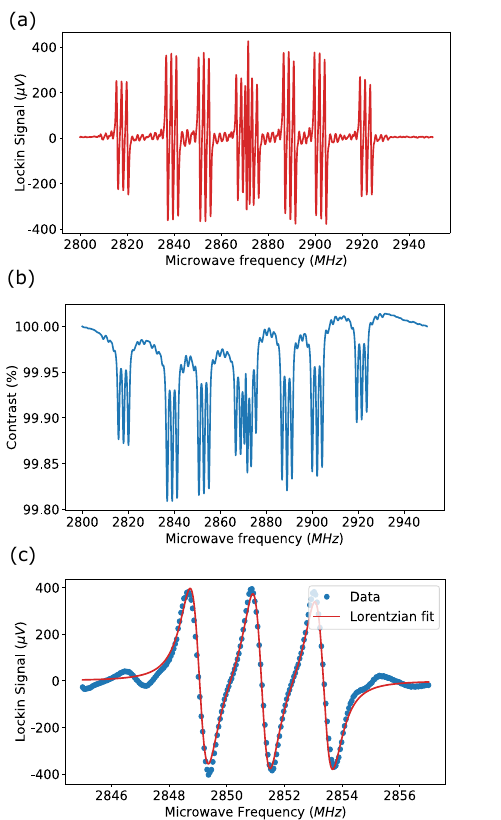}%
 \caption{\label{fig: ODMR}(a) FM ODMR spectrum of the demodulated signal. Four pairs of resonances, each 
 corresponds to $m_s =0$ to $m_s=\pm1$ transition of one of the crystallographic axes of NV center ensemble 
 including the three hyperfine transitions due to $^{14}$N nuclear spins. The spectrum is obtained at 
 \SI{1}{\kilo \hertz} modulation frequency, -3 dBm MW power at signal source, \SI{400}{\kilo \hertz} 
 maximum frequency deviation, \SI{10}{\ms} 
 lockin time constant, and \SI{700}{\hertz} input high pass filter frequency. (b) Normalized integrated 
 ODMR spectrum of 
 the spectrum in (a) (for details, see supplementary information \RomanNumeralCaps{1}), (c) Zoomed-in ODMR spectrum corresponding to resonance peak in the range  $2.845-\SI{2.860}{\giga \hertz}$. The extracted linewidth of each hyperfine level is $\sim \SI{1}{\mega \hertz}$}.
 \end{figure}

The optically detected magnetic resonance (ODMR) spectrum is obtained by sweeping the MW frequency and 
observing the in-phase quadrature as illustrated in Fig. \ref{fig: ODMR}(a), the signal in the other 
quadrature is minimized by adjusting the phase difference between modulation and demodulation. The sensitivity of the magnetometer characterizes the minimum detectable magnetic field after an integration time of \SI{1}{\second}, and is given by\cite{doi:10.1073/pnas.1601513113}:
\begin{equation}
\eta_{esr}= \frac{\sigma \sqrt{\tau}}{\gamma_{e}\frac{dV_{lockin}}{df}|_{max}},
\label{eqn: exp sensitivity}
\end{equation}
where $\sigma$ is the standard deviation of the ODMR spectrum, $\gamma_e$ is the electron spin gyromagnetic ratio, $\frac{dV_{lockin}}{df}|_{max}$ is the maximum slope, and $\tau$ is the integration time (for detailed calculation see supplementary information \RomanNumeralCaps{1}). The maximum slope is calculated by fitting the ODMR curve with the derivative of a Lorentzian function [Fig. \ref{fig: ODMR}(c)] and finding the point where the slope of the derivative Lorentzian is maximum. For the ODMR spectrum in Fig. \ref{fig: ODMR}(c) the maximum slope is \SI{2.20}{\nano \volt \per \hertz}, $\sigma$ is $ 15 \pm \SI{3}{\micro \volt}$, integration time is \SI{10}{\ms}, and the sensitivity is $10 \pm \SI{3.5}{\nano \tesla \per \sqrt \hertz}$. The shot-noise limited sensitivity is given by\cite{PhysRevB.84.195204}:
\begin{equation}
\eta_{shot-noise}= P_{F}\frac{1}{\gamma_e} \frac{\Delta\nu}{C\sqrt{R}},
\label{eqn: shotnoise sensitivity}
\end{equation}
where $P_F$ is the numerical factor resulting from the line shape of the resonance,  $\Delta\nu$ is the linewidth, $C$  the ODMR 
contrast, and $R$ is the  photon detection rate. For Lorentzian 
line shape, $P_F=4/3\sqrt{3}$. For out ODMR spectrum, contrast is 0.15\%, linewidth is \SI{1.0}{\mega 
\hertz}, photon rate $R$ is \SI{7.5e14}{photons \per \sec}, and  the photon shot-noise limited sensitivity is 
\SI{0.7}{\nano \tesla \per \sqrt \hertz}.

\begin{figure*}
 \includegraphics[width=\textwidth]{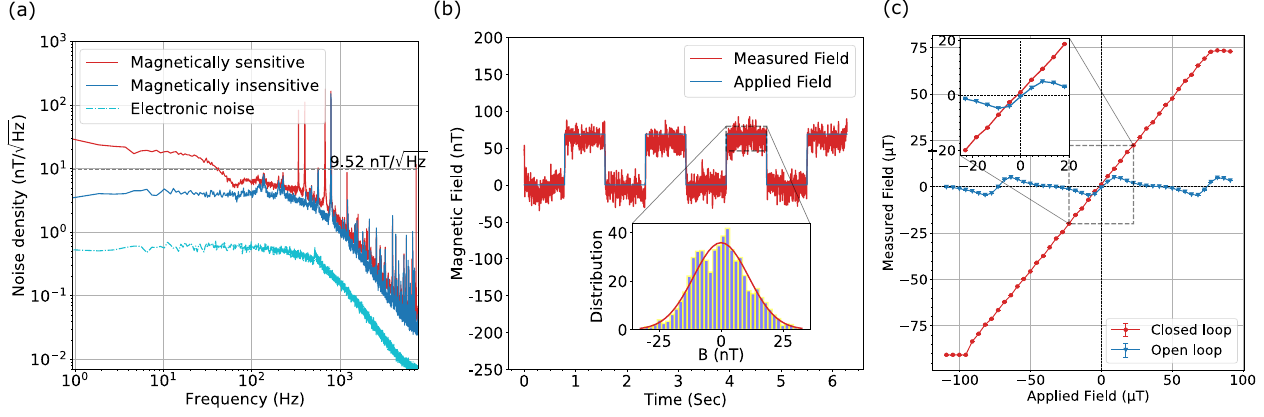} %
 \caption{\label{fig: Tracking and Dynamicrange} (a) Noise spectral density plot. Fifty time-traces 
 were acquired at \SI{1.9}{\kilo samples \per \sec}, \SI{10}{\milli \second} lockin integration time, \SI{1}{\kilo \hertz} modulation frequency, and $f_{dev}$ of \SI{400}{\kilo \hertz}, for magnetically sensitive, insensitive, and laser off case. (b) Magnetic field tracking with square-wave test field. The 
 MW frequency is set to resonance frequency. The time-traces are acquired at \SI{1.9}{\kilo samples \per \sec}, \SI{25}{\milli \second} lockin integration time, \SI{1}{\kilo \hertz} modulation frequency, and frequency deviation of \SI{400}{\kilo \hertz}. Inset 
 figure shows the histogram and the fitted Gaussian distribution  of the  fluctuations for one second duration. 
 (c) Dynamic range comparison of the closed-loop and the open-loop measurement.}%
 \end{figure*}

The performance of the magnetometer is further analyzed with noise power spectral density (PSD).
The PSD plots are shown in Fig, \ref{fig: Tracking and Dynamicrange}(a). PSD plots are obtained by setting the 
microwave frequency at resonance (magnetically sensitive), away from resonance (magnetically insensitive), and without laser excitation (electronic noise). The noise 
power is given by:
\begin{equation}
\sigma^2 = \int_{-\infty}^{\infty} |x(t)|^2 dt=\int_{-\infty}^{\infty}|X(f)|^2df,
\label{eqn: noise power}
\end{equation}
where $\sigma$ is the standard deviation,  $x(t)$ is the signal,  and $X(f)$ is its Fourier transform.
 \begin{figure}
 \includegraphics[width=0.5\textwidth]{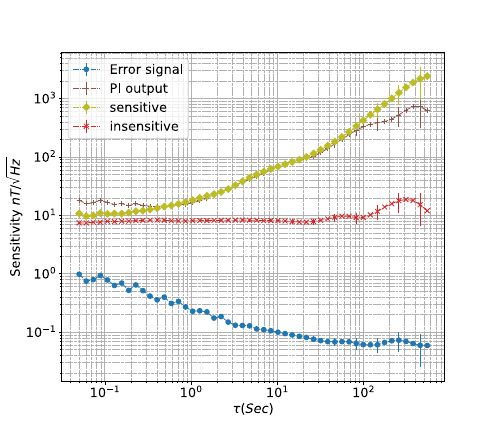}
 \caption{\label{fig: AllanDeviation}Allan deviation measurement. The confidence intervals are calculated 
 by the $\chi^2$ statistics \cite{riley2008handbook}. Olive-colored plot indicates open-loop 
 magnetically sensitive, red-colored magnetically insensitive, 
 brown color closed-loop PI output, and blue color is the error in the PI output. The proportional 
 and integral gains for the closed-loop case are -50 and -2.5, respectively.}%
 \end{figure}

The magnetic noise spectral density is obtained by taking the fast Fourier transform (FFT) of time 
trace data acquired and normalized (see supplementary information \RomanNumeralCaps{2}), scaled by the 
zero-crossing slope (ZCS), and the 
gyromagnetic ratio of the ODMR curve. Fifty  time traces each for one second were acquired for 
magnetically sensitive case, magnetically insensitive case, and  laser excitation turned off to acquire 
the electronic noise 
floor. The mean noise floor within  \SI{10}-\SI{100}{Hz} of the magnetometer is \SI{9.52}{\nano \tesla \per \sqrt \hertz}.

The sensitivity of the magnetometer is also determined by applying a square-wave test magnetic field as 
illustrated in Fig. \ref{fig: Tracking and Dynamicrange}(b). A 100-turn current-carrying coil 
(\SI{5}{\cm}) is placed in 
the vicinity of the magnetometer, and the coil field is calibrated by sweeping the DC current in the 
test 
coil and observing the shift in the resonant frequency of the ODMR spectrum. The calibration constant 
of the 
coil is derived by the linear fitting of the frequency shift vs. DC current and extracting the slope 
of the 
linear fit (see supplementary information \RomanNumeralCaps{3}). The extracted slope is \SI{137}{\kilo \hertz \per \ampere}, which translates to \SI{4.9}{\micro 
\tesla \per \ampere} when scaled by the gyromagnetic ratio. The coil is driven with a modulated 
current, generating a magnetic field of $\sim$ \SI{50}{\nano \tesla}, and the MW frequency is set to 
the ZC point of the ODMR spectra. Hundred synchronous eight-second time traces were acquired at the 
sampling rate of \SI{1.9}{\kilo samples \per \sec} and offset-corrected and averaged in time domain. 
The fluctuation in the detected magnetic  for a \SI{1}{\sec} window is shown in the inset of Fig. 
\ref{fig: Tracking and 
Dynamicrange}(a). The standard deviation of the fluctuation is \SI{11}{\nano \tesla}, which translates 
to a sensitivity of 
\SI{12}{\nano \tesla \per \sqrt \hertz}.

The linear dynamic range of the magnetometer for the open loop and the 
closed-loop\,\cite{10.1063/1.5034216} cases is compared in 
Fig. \ref{fig: Tracking and Dynamicrange}(c). The MW frequency is set to resonance frequency for the 
closed-loop case, 
and the lockin output is sent to the digital signal processing based PI controller implemented on the 
FPGA.
The PI output signal generates an error signal to shift the frequency of the MW signal generator to 
track the resonance frequency. The dynamic range for closed-loop case 
in the order of $\sim$\SI{200}{\micro \tesla} compared to open-loop case of $\sim$\SI{10}{\micro 
\tesla}, a \SI{20}{\times} improvement in the dynamic range without compromising the sensitivity. The 
dynamic range is 
limited by the maximum frequency deviation ($f_{dev}$) possible with the MW signal generator, in this
case, 
\SI{5}
{\mega \hertz}. The dynamic range can be pushed further by using the signal 
generators with higher $f_{dev}$.

The sensor performance was further analyzed with Allan deviation measurements. The previous methods 
(eq. (\ref{eqn: exp sensitivity}) and Fig. \ref{fig: Tracking and Dynamicrange}(a,b)) to calculate the 
sensitivity assume that the noise is white within the integration time. But, for longer averaging 
time, diverging noise sources are present due to the dependence of experimental parameters on ambient 
conditions. 

The overlapping Allan deviation measurement\,\cite{riley2008handbook} was performed for three 
different cases: open-loop magnetically sensitive case, oper-loop magnetically insensitive case, and 
the closed loop case. The signal was recorded for a \SI{10}{\minute}  duration as illustrated in Fig. 
\ref{fig: AllanDeviation}. The Allan deviation, $\sigma(\tau)$, times the integration time 
$\sqrt{\tau}$ gives the sensitivity at given $\tau$.
The magnetically insensitive measurement represents the drifts due to the laser source, ($\sigma(\tau) 
\times \sqrt{\tau}$) is flat over \SI{100}{\second} integration time, which implies noise is white in 
frequency\,\cite{riley2008handbook}. In magnetically sensitive measurement, the flat portion of the plot 
is $\approx \SI{10}{\nano \tesla \per \sqrt \hertz}$ confirming the sensitivity obtained from previous 
measurements. At larger $\tau$, $\sigma(\tau)$ deviates due to the dependence of  resonance frequency 
on 
temperature (\SI{\approx74}{\kilo \hertz \per \deg \kelvin})\,\cite{schirhagl2014nitrogen}. The Allan deviation measurement was also 
performed for the closed-loop feedback case, which overlaps with the magnetically sensitive open-loop case. The error signal from the PI controller has $-\frac{1}{2}$ slope, showing that the signal has a white phase 
noise\cite{riley2008handbook}, implying that the drifts are transferred to the PI output to reduce the 
error in the lockin quadrature.
\begin{figure}
 \includegraphics[width=0.5\textwidth]{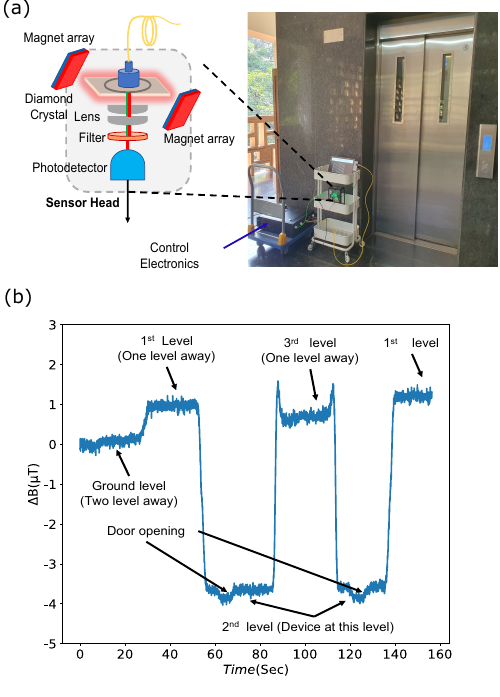}
 \caption{\label{fig: Elevator field tracking} (a) Photograph of magnetometer placed \SI{60}{\cm} away from the door of the elevator on the second level of the four-story building along with the control electronics box. Inset figure shows the schematic of the sensor head with magnet array used to provide the bias field (b) Temporal response of the magnetometer when the elevator moves from ground level to third level of the building with the magnetometer placed at second level. The data is acquired at the sampling rate of \SI{1.9}{\kilo samples \per \sec}, and \SI{40}{\milli \second} lockin integration time. A Gaussian filter with $\sigma = 50$ is applied to raw data.}%
\end{figure}

As an application, the magnetometer was utilized to detect the real-time magnetic field generated by the movement of an elevator from the ground level to the third level and the door opening. The magnetometer was placed on the second level and \SI{60}{\cm} away from the elevator door of a four-story building (each level is \SI{3.35}{\meter} above the other, and the elevator door height is \SI{2.28}{\meter}). A photograph of the device is shown in Fig. \ref{fig: Elevator field tracking}(a), and the inset shows the schematic of the sensor head. The device was operated in closed-loop feedback control mode, and the PI controller generates a PI signal to lock the magnetometer output at the maximally sensitive resonance point in response to the external magnetic field. Fig. \ref{fig: Elevator field tracking}(b) illustrates the response of the magnetometer, as the elevator moves to the first level from the ground level. The change in magnetic field is observed in real time. When the elevator reaches the second level, the largest change in the magnetic field is observed because the elevator car is nearest to the device. Also, a change is observed when the door opens due to the magnetic field generated by the door opening mechanism. Upon reaching the third level, the response returned to the same level as the first level since the distance from the device was approximately the same. A similar response is observed when the elevator returns to the first level from the third level.

In conclusion, we have realized a novel portable magnetometer with \SI{\sim 10}{\nano \tesla \per \sqrt \hertz} sensitivity 
while at the same time extending the dynamic range beyond the intrinsic range by locking the frequency to the NV resonance that can be operated under normal temperature and humidity conditions. We characterized the magnetometer for metrics such as 
magnetic noise density, sensitivity, linear dynamic range, and stability. The sensitivity can be further 
improved by engineering the diamond, laser noise cancellation, and efficient collection of photons emitted 
from the diamond. Our work paves the way for applying the portable magnetometer in a wide variety of industrial applications in the near future.

\section*{acknowledgment}
K.S. acknowledges financial support from DST Quest Grant DST/ICPS/QuST/Theme-2/2019/Q-58, IIT Bombay Seed Grant 17IRCCSG009 and SERB power research grant SPG/2023/000063. 
The authors acknowledge support from staff and access to facilities at the Wadhwani Electronics Lab (WEL), Department of Electrical Engineering, IIT Bombay for building the system reported in this work. 
H.K acknowledges funding from the Prime Minister's Research Fellowship. The authors acknowledge insightful discussions with Dr. Madhur Parashar, IIT Bombay, and Prof. Saikat Ghosh, IIT Kanpur.



%
%

%


\bibliography{references}

\end{document}